\documentclass{pasj00}
\draft

\newcommand{\Dt}{\Delta t}
\SetRunningHead{R.-F. Shen and L.-M. Song}{Variability Time Scales of GRBs}

\title{Characteristic Variability Time Scales of Long Gamma-Ray Bursts}

\author{Rong-Feng \textsc{Shen} and Li-Ming \textsc{Song}}
\affil{Particle Astrophysics Center, Institute of High Energy Physics, \\
Chinese Academy of Sciences, Beijing\, 100039, P. R. China}
\email{shenrf@mail.ihep.ac.cn, songlm@mail.ihep.ac.cn}
\Received{$\langle$2002 December 4$\rangle$}
\Accepted{$\langle$2003 January 14$\rangle$}


\begin{document}
\maketitle

\begin{abstract}
We determined the characteristic variability time scales ($\Dt_{\rm p}$) 
of 410 
bright and long GRBs, by locating the peaks of their Power Density 
Spectra, defined and calculated in the time domain. We found that the 
averaged variability time scale decreases with the peak flux. This is 
consistent with the time-dilation effect expected for the cosmological 
origin of GRBs. We also found that the occurrence distribution of the 
characteristic variability time scale shows bimodality, which might be 
interpreted as that the long GRB sample is composed of two sub-classes 
with different variability time scales. However, we found no difference 
for some other characteristics of these two sub-classes.  
\end{abstract}

\section{Introduction}

The isotropy of gamma-ray burst (GRB) directions and the deficiency of 
weak bursts suggest a 
cosmological origin of GRBs \citep{meegan92}, which has been confirmed 
by the detection 
of GRB afterglows \citep{costa97,van97,metzger97,kulkarni98}. 
An energy release of $\sim 10^{51}-10^{53}$ erg (assuming no beaming), implied
by the cosmological origin, makes them to be outstanding events in the universe.

According to the most popular ``fireball'' model \citep{piran99}, the observed 
$\gamma$-rays are emitted when an ultra-relativistic flow is converted to 
radiation. It has been suggested that the energy conversion occurs either 
due to an interaction with the external medium (``external shocks'') 
\citep{rees92} or due to collisions 
within the flow (``internal shocks'') \citep{rees94}.

The progenitors of GRBs (``central engine'') remain the most mysterious aspects,
since 
they are hidden from direct observations. The most popular models are mergers
of two compact objects (NS -- NS merger or NS -- BH merger), or the collapse of massive
stars. Currently, it is difficult to distinguish them from observations. 
It is also likely that 
more than one type of progenitor could give rise to GRBs, and the overall 
GRBs may comprise different sub-classes corresponding to different progenitors.
The most well-known classes are short and long classes separated at
$T_{90} \sim$ 2 s, arising from their bimodal duration ($T_{90}$) distribution 
\citep{kouve93}. In recent years, a few studies have been devoted to 
a reclassification of GRBs using multivariate analysis, which have yielded three 
sub-classes \citep{mukherjee98,bala01}. 
  
About 2000 GRBs \footnote{$\langle$
http://gammaray.msfc.nasa.gov/batse/grb/catalog/current/ 
$\rangle$.} 
have been detected and recorded by the Burst
And Transient Source Experiment (BATSE) on board the 
Compton Gamma-Ray Observatory.
They have complicated and irregular time profiles which vary drastically
from one burst to another. Recent studies concerning the internal shock paradigm show 
that the observed complicated temporal structures are directly associated 
with the activity of the central engine \citep{koba97}. 
The observed variability provides an intersting clue as to the nature of GRBs
\citep{piran99}. 

Observationally, the light curve of a burst consists of successive pulses, 
which may be the fundamental emission units. Many people have decomposed 
light curves into pulses using model-dependent fitting \citep{norris96} 
or peak finding selection \citep{nakar02,mcbreen01} and
analysed the temporal properties of the resulting pulses, such as the rise-time, 
decay time, FWHM, and time interval between the neighbouring pulses. However,  
the real time profiles are so irregular that many pulses are overlapping and 
unseparable, prohibiting a clear decomposition of those pulses.
 
In this work we adopted a different approach. We calculated 
the power density spectra (PDS) in the time domain for 410 bright long 
bursts, and regarded the time scale corresponding to the maximum of the PDS 
as the time scale of typical variations in the profile, and thus defined 
it as the characteristic variability time-scale of the burst. We then studied
the distribution of the characteristic variability time scale and the
correlation between the time scale with the GRB intensity.
     
\section{PDS in the Time Domain}

Many authors have calculated the PDSs of GRBs based on the Fourier 
transformation (\cite{belo98}, 2000; \cite{pozanenko00}). 
However, Fourier analysis cannot replace direct variability study 
in the time domain. Except for the periodic and quasi-periodic processes, 
there is no direct 
correspondence between a structure in the Fourier spectrum and the physical 
process taking place at a certain time scale. The power density at 
a given Fourier frequency 
can result from contributions from different processes on different time scales.

Without using the Fourier transformation, a new technique for timing 
analysis in the time domain has recently been proposed \citep{li01,li02}. 
Quantities characterizing the temporal properties, e.g., power density, 
coherence, and time lag, can be defined and calculated directly in the 
time domain with this technique. 

Following \citet{li01}, the variation power in a light curve, $x(k)$, 
is defined as
\begin{equation}
 P(\Dt)=\frac{Var(x)}{(\Dt)^2}=
\frac{{\displaystyle \frac{1}{N}}\sum_{k=1}^{N}[x(k)-\bar{x}]^2}{(\Dt)^2} ,   
\end{equation}
where $x(k),k=1,...,N$, is a counting series obtained from the time history of
the observed photons with a time step of $\Dt$; $\bar{x}=\sum_{k=1}^{N}x(k)/N$ is 
the average counts. The power, $P(\Dt)$, represents the contribution from 
the variation at all time scales $\geq \Dt$. The power density, $p(\Dt)$, 
can be defined in the time domain as the rate of change of $P(\Dt)$ 
with respect to the time step, $\Dt$. From two powers, $P(\Dt_1)$ and $P(\Dt_2)$, 
at two time scales, $\Dt_1$ and $\Dt_2$ ($\Dt_2 > \Dt_1$), the power density at 
time scale $\Dt=(\Dt_1+\Dt_2)/2$ is evaluated approximately by
\begin{equation}
p(\Dt)=\frac{P(\Dt_1)-P(\Dt_2)}{\Dt_2-\Dt_ 1} 
\end{equation}
For a noise series where $x(k)$ follows the Poisson distribution, 
the noise power is given by
\begin{equation}
P_{\rm noise}(\Dt)=\frac{Var(x)}{(\Dt)^2}=\frac{\langle x \rangle}{(\Dt)^2} ,
\end{equation}
and the noise power density at $\Dt=(\Dt_1+\Dt_2)/2$ is
\begin{equation}
p_{\rm noise}(\Dt)=\frac{P_{\rm noise}(\Dt_1)-P_{\rm noise}(\Dt_2)}{\Dt_2-\Dt_1} .
\end{equation}
The signal power density can then be defined as 
\begin{equation}
p_{\rm signal}(\Dt)=p(\Dt)-p_{\rm noise}(\Dt) .
\end{equation}
 
\section{Data Reduction}

We used the BATSE Concatenated 64-ms Burst Data summed over energy 
channels II and III (50 --- 300 keV), and we 
utilized the results of a background fitting for 64-ms Burst Data \footnote{
$\langle$ 
http://cossc.gsfc.nasa.gov\-/ba\-ts\-eb\-ur\-st/s\-ixtyfour\_ms/bc\-kgn\-d\_f\-it\-s.html
$\rangle$.}.
In order to reduce the influence of the background, one would like to cut the light 
curve with a window neatly including the whole burst. We defined and used the 
``$T_{100}$ window'', where $T_{100}=T_{90}/90\%$. The window was obtained by
extending the $T_{90}$ window forward for $0.05T_{90}$ and backward for 
$0.05T_{90}$, respectively. 
   
Since we are interested in using the PDS to investigate the variation power 
distribution over a large range of time scales, short bursts are not suitable 
for calculating the PDS, except that their durations are long compared with 
the timing resolution. We thus selected only bursts with $T_{100} > 15$ s. 
To avoid large Poisson fluctuations in the light curve, we excluded dim bursts 
with peak count rates $<$ 250 counts per 64 ms bin. These two criteria gave 
a sample of 478 bursts from the BATSE Current Catalog 
\footnote{$\langle$
http://gammaray.msfc.nasa.gov/batse/grb/catalog/current/ $\rangle$.}. 
We then calculated the PDS of each burst in the time 
domain with the algorithm described in section 2. 

\section{PDS Calculation}

For an individual burst, the power densities are calculated at time 
scales of $\Dt$, arranged at equal intervals in logarithmic space. 
To avoid fluctuations in the PDS, we binned PDS with an approximately 
equal number of data points in each time-scale bin, and the uncertainties 
of the data were derived from the binning. 

Most of the bursts' PDSs showed a ``bump'' shape. The bump-shape indicates
variation peaks at a specific time scale; we regard the peak time scale, 
$\Dt_{\rm p}$, as being the characteristic variability time scale of the burst.

\begin{figure}
  \begin{center}
    \FigureFile(50mm,60mm){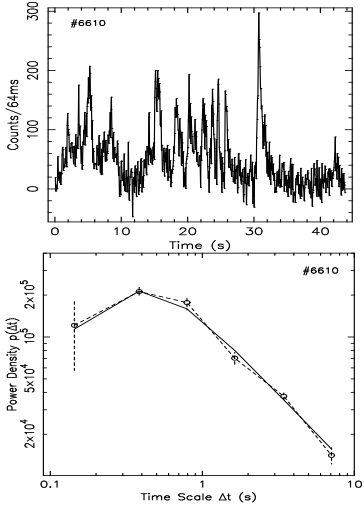}
  \end{center}
  \caption{Time profiles of a burst (upper panel) and its PDS
   calculated in the time domain (lower panel). The solid line is the 
   broken power-law model fit to the calculated PDS.}\label{fig:6610}
\end{figure}

To improve the accuracy of $\Dt_{\rm p}$, we fitted the PDS with a 
broken-power-law model, 
$p(\Dt)=
{\displaystyle \frac{(\Dt/P_1)^{P_2+P_3}}{(\Dt/P_1)^{P_2}+(\Dt/P_1)^{P_3}}}P_4$;
the fit was acceptable for most bursts. 
The characteristic variability time scale $\Dt_{\rm p}$ was determined from the peak 
location of the fitted PDS curve. Figure~\ref{fig:6610} gives an example 
of a burst's time profile and its PDS.

\section{Characteristic Variability Time Scales}

\subsection{$\Dt_{\rm p}$ for a Simulated Time Series}

Does $\Dt_{\rm p}$ really reflect the variability time scale of the typical 
variation? How is $\Dt_{\rm p}$ correlated with the sizes of pulses if we 
consider that a burst is composed of stochastic pulses?
To answer these questions, we simulated a long stochastic time series 
with a 16 ms time resolution and 4800 s duration. The time series were 
divided into 100 segments, and the power densities at different time scales 
were calculated for each segment. The resulting PDS was obtained by 
averaging the PDSs over 100 segments. The pulse model was taken from 
\citet{norris96}, with the following pulse shape: 
\begin{eqnarray} 
I(t) & = & A\,\,{\rm exp} [-(|t-t_{\rm max}|/\tau_{\rm r})^\nu], \ \ \ t<t_{\rm max}, \\
     & = & A\,\,{\rm exp} [-(|t-t_{\rm max}|/\tau_{\rm d})^\nu], \ \ \ t>t_{\rm max}, 
\end{eqnarray}
where $t_{\rm max}$ is the time of the pulse's peak, $A$ is the pulse amplitude, 
$\nu$ is 
the ``peakedness'' parameter, and $\tau_{\rm r}$ and $\tau_{\rm d}$ are the rise and 
decay time constants, respectively. 

\begin{figure}
  \begin{center}
    \FigureFile(60mm,40mm){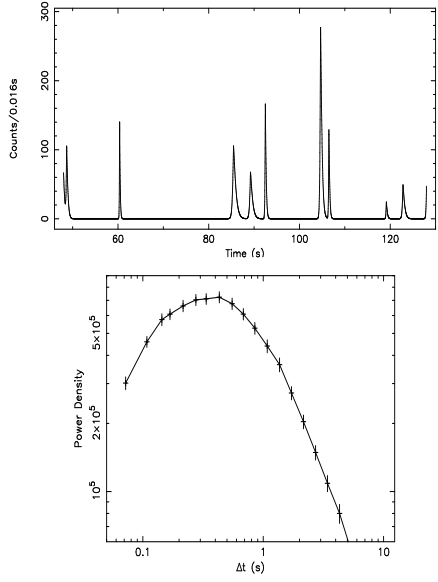}
  \end{center}
  \caption{Simulated time series (upper panel) and 
   the calculated PDS (lower panel). The stochastic 
   pulse parameters are $\langle {\rm ln} (\tau_{\rm r}) \rangle = -2.0$, 
   $\tau_{\rm d}/\tau_{\rm r}$=2.5.}
\label{fig:simu}
\end{figure}

\begin{figure}
  \begin{center}
    \FigureFile(60mm,50mm){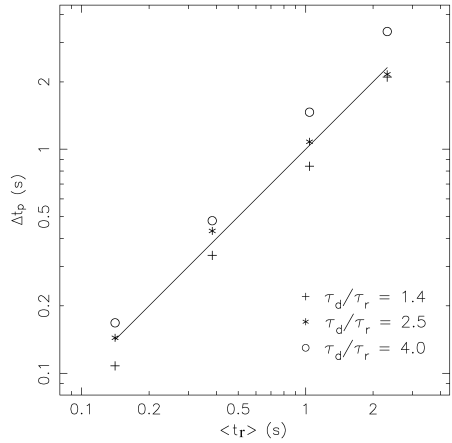}
  \end{center}
  \caption{Characteristic variability time scales $\Dt_{\rm p}$ vs. 
   $\langle t_{\rm r} \rangle$, for the simulated time series. See 
   definition of $t_{\rm r}$ in the text. The solid line represents 
   $\Dt_{\rm p} = \langle t_{\rm r} \rangle$. Note the correlation between 
   $\Dt_{\rm p}$ and the average rising time, $\langle t_{\rm r} \rangle$, 
   of the stochastic pulses.}
\label{fig:tpvstr}
\end{figure}

We assumed that the interval between the neighboring pulses is 
exponentially distributed with a mean value of 10 s, $A$ is uniformly
distributed within 10 --- 200 counts (0.016s)$^{-1}$, $\nu = 1.2$, the rise-time
constants, $\tau_{\rm r}$, are a Log-Normal distribution with a standard deviation
$\sigma({\rm ln} \ \tau_{\rm r}) = 0.5$, and the decay-to-rise ratio, 
$\tau_{\rm d}/\tau_{\rm r}$, is a 
constant. We defined the pulse rising time, $t_{\rm r}$, as the time during
which the intensity, $I(t)$, increases from $5\%A$ to the peak.
We then chose different values for the mean 
$\langle {\rm ln} \ \tau_{\rm r} \rangle$ 
and the decay-to-rise ratio, $\tau_{\rm d}/\tau_{\rm r}$, to simulate the time 
series, and to calculate the PDS in the time domain. 
The characteristic variability 
time scale, $\Dt_{\rm p}$, was determined from the peak of the PDS. 
Figure~\ref{fig:simu} shows a simulated time series and the calculated 
PDS, respectively. Figure~\ref{fig:tpvstr} plots $\Dt_{\rm p}$ vs. 
$\langle t_{\rm r} \rangle$. Figure~\ref{fig:tpvstr} shows that $\Dt_{\rm p}$ 
is truly 
the variability time scale of typical variations, and is sharply correlated 
with the rising time, $t_{\rm r}$, of the typical pulses. 	 

\subsection{Distribution of $\Dt_{\rm p}$}

\begin{figure}
  \begin{center}
    \FigureFile(60mm,60mm){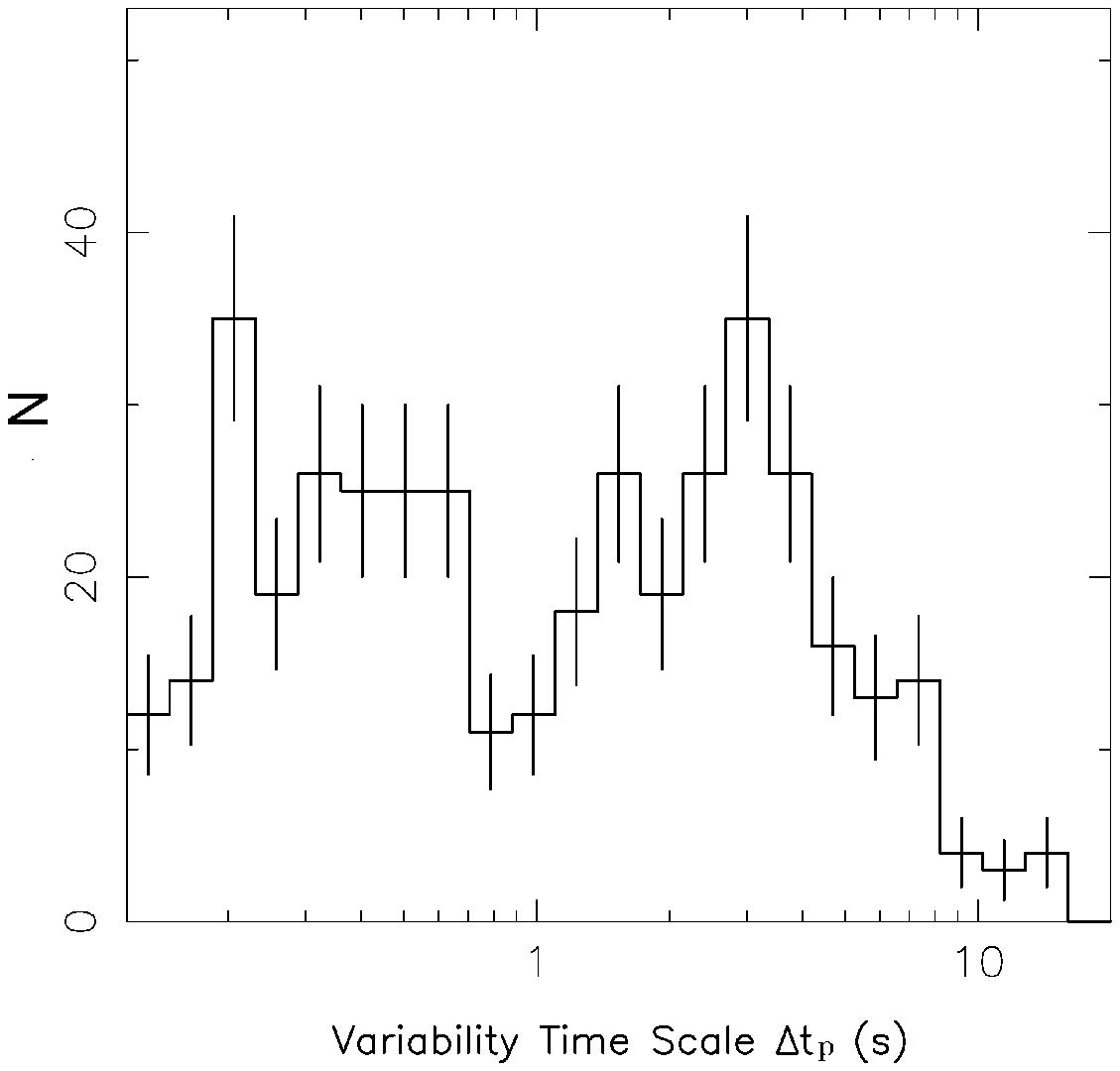}
  \end{center}
  \caption{Histogram distribution of the characteristic variability 
   time scales for 410 bright long bursts.}
\label{fig:occur}
\end{figure}

Among the PDS samples, 63 PDSs keep rising untill the smaller 
time-scale limit where the PDS can be calculated, without showing any
decrease. 
This means that the characteristic time scales of those bursts are smaller 
than, or equal to, the smaller time-scale limit, whereas they cannot yet be 
determined. After discarding the 63 samples and 5 badly fitted 
($\chi^2/\nu > 50$) samples, we obtained $\Dt_{\rm p}$ for 410 bursts. 
Figure~\ref{fig:occur} plots their histogram distribution; it is a bimodal 
distribution with a the demarcation at $\Dt_{\rm p} \sim 1$ s. 

To investigate the reliability of the bimodality, we performed a 
chi-squared test \citep{press92} for the differences between 
the distribution data and a known uniform distribution in the range of 
$\Dt_{\rm p} \sim$ 0.1 --- 8.0 s, where 19 bins were included. The mean of the 
uniform distribution was set to that of the data sets. The test gave a 
chi-squared probability, $Q(\chi^2|\nu)=1.437\times10^{-4}$, 
where the small value of $Q(\chi^2|\nu)$ indicates a significant difference 
between these two distributions.
The bimodal distribution of $\Dt_{\rm p}$ indicates that bursts of 
$\Dt_{\rm p} >$ 1 s 
and $\Dt_{\rm p} <$ 1 s may belong to two GRB sub-classes respectively.

\subsection{Time Dilation Test}

If a GRB occurs at a cosmological distance, then every structure in the 
time profile will be stretched by a factor $1+z$ due to the expanding 
universe, where $z$ is the red shift. Therefore, we should observe that the 
dimmer bursts have larger characteristic variability time scales than do the
brighter bursts, assuming that bursts at different cosmological distances are 
``standard candles'' with the same intrinsic characteristic variability 
time scale.

\begin{figure}
  \begin{center}
    \FigureFile(60mm,60mm){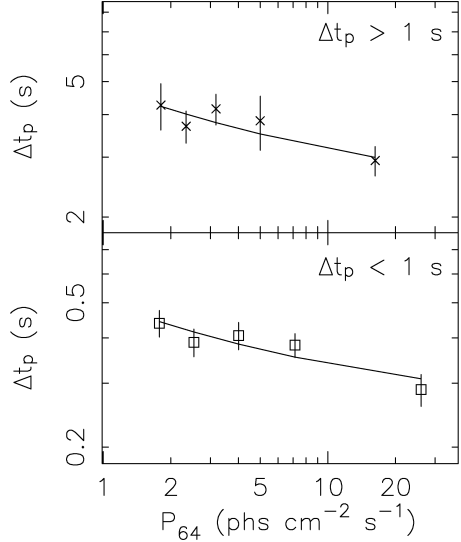}
  \end{center}
  \caption{$\Dt_{\rm p}$ vs. brightness distribution for the slow variable 
   group ($\Dt_{\rm p} > 1$ s) and the fast variable group 
   ($\Dt_{\rm p} < 1$ s). Each bin 
   includes an equal number of samples. The solid lines are the best-fit model 
   predictions.}
\label{fig:tpvsp64}
\end{figure}

According to the bimodal distribution of $\Dt_{\rm p}$, we divided the bursts 
into the $\Dt_{\rm p} >$ 1 s group and the $\Dt_{\rm p} <$ 1 s group; 
figure~\ref{fig:tpvsp64} is a plot of the distribution of the mean of 
$\Dt_{\rm p}$ in 5 
brightness bins for each group, where the brightness is represented by 
$P_{64}$, the peak flux measured at the 64 ms time scale. Both groups 
show that the averaged $\Dt_{\rm p}$ decreases with the brightness, 
which is consistent 
with the cosmological origin of GRBs.   

\begin{table}
\begin{minipage}{14cm}
\caption{Fitted results for time dilation. \footnote{
The estimated uncertainties of $\Dt_0$ and the luminosity 
correspond to the 90\% confidence level.}}\label{tab}
\begin{tabular}{ccccc}
\hline\hline
Group & $\Delta t_0$ (s) & Peak luminosity ($10^{50}$erg s$^{-1}$) & 
			$z_{\rm min}$ & $z_{\rm max}$	\\
\hline
$\Dt_{\rm p} >$ 1 s & $2.30(\pm 0.18)$ & $7.56(\pm 2.81)$ & 
			$0.28(\pm 0.16)$ & $0.85(\pm 0.32)$	\\
$\Dt_{\rm p} <$ 1 s & $0.256(\pm 0.004)$ & $5.52(\pm 0.36)$ & 
			$0.13(\pm 0.11)$ & $0.72(\pm 0.15)$	 \\
\hline
\end{tabular} 
\end{minipage}
\end{table} 

To quantitatively test the time-dilation effect, we fitted the 
$\Dt_{\rm p}$ vs. brightness distributions with a model prediction; the model
is described in the Appendix. Table~\ref{tab} lists the fitting results for 
the $\Dt_{\rm p} > 1$ s group and the $\Dt_{\rm p} <$ 1 s group, respectively, 
in a standard 
$\Omega=1, \Lambda=0$ cosmology, with the normalized Hubble constant, 
$h$, set to 0.75. 
The luminosity is the product of the peak photon number luminosity 
and the assumed mean photon energy, 150 keV. The time dilation factors, 
$(1+z_{\rm max})/(1+z_{\rm min})$, derived for the $\Dt_{\rm p} > 1$ s group and 
the $\Dt_{\rm p} <$ 1 s group are $1.46(\pm 0.27)$ and $1.52(\pm 0.20)$, 
respectively.

\section{Discussion}

We calculated the PDSs of 410 bright long bursts in the time domain, and 
determined their characteristic variability time scales by locating the 
peaks of their PDSs. The distribution of $\Dt_{\rm p}$ is a bimodal 
distribution with the demarcation at $\Dt_{\rm p} \sim 1$ s. GRBs may be divided 
naturally into two groups, a fast variable group and a slowly variable 
group.

Following \citet{koba97} and \citet{piran99}, in the internal shock 
paradigm, the emitted radiation from each collision between two 
relativistic shells will be observed as a single pulse, whose time 
scale depends on the cooling time, the hydrodynamic time, and the 
angular spreading time. In most of cases, the electron cooling time 
is much shorter than the latter two. The hydrodynamic time scale is 
determined by the time that the shock crosses the shell, whose width
is $d$. A calculation revealed that this time scale (in the observer's 
rest frame) is of the order of the light crossing time of 
the shell, $d/c$. If the collisions of the shells take place 
at a larger radius, $R$, angular spreading \citep{sari97} affects the 
time scale of the pulse. If the distance between shells is $\delta$, 
the resulting angular spreading time for the pulse is $\sim \delta/c$. 
If $\delta > d$, the observed variability time scale will be determined 
by the angular spreading. In any case, the observed temporal structure 
is directly associated with the variability of the ``central engine'', 
and the observed time scale is proportional to the size of the ejected 
wind or the separation between the winds. Consequently, it is likely 
that the observed time scale is correlated with the size of the 
progenitor.  

Our results concerning the existence of a fast variable group and a slowly 
variable group may indicate that the bright long bursts comprise two 
sub-classes generated from different sizes of ejected winds or 
distances between winds, or from different sizes of the progenitors.
Both groups show the trend that the characteristic variability 
time scale decreases with the brightness of the burst, consistent 
with the cosmological time-dilation effect. We fitted the data to
a model for both the fast variable group and the slowly variable 
group, respectively. The smaller values of $z_{\rm max}$ and the time-dilation 
factors, compared with previous results \citep{norris94,che97},
may be because our sample does not include more weak bursts.

\section*{}
The authors thank S. N. Zhang, T. P. Li, W. F. Yu, F. J. Lu and J. L. Qu
for their useful discussions and valuable advice. We thank S. N. Zhang
for carefully reading the manuscript. This work was supported
by Special Funds for Major State Basic Research Projects of China,
and has made use of data obtained from the High Energy Astrophysics 
Science Archive Research Center (HEASARC) provided by NASA's Goddard 
Space Flight Center.

\appendix
\section*{Model Prediction for Time Dilation}

The cosmological time-dilation effect will stretch every time structure of 
GRBs if they are at cosmological distances. Assuming that GRBs have the same 
characteristic variability time scale, $\Dt_0$, we should observe 
\begin{equation} 
\Dt=\Dt_0(1+z) .
\label{td}
\end{equation} 

Since the red shift, $z$, is unknown, the observed peak flux, $P$ 
[phs s$^{-1}$ cm$^{-2}$], is used to express $z$ under the 
``standard candle'' assumption. For simplicity, we introduce the peak 
photon number luminosity, $L_{\rm n}$ [phs s$^{-1}$], as the standard luminosity 
of GRBs, without considering the photon energy redshift. For a standard 
Friedmann cosmology ($\Omega=1, \Lambda=0$), the luminosity distance is 
\begin{equation} 
d_L=2R_0 (1+z-\sqrt{1+z}) , 
\label{ld}
\end{equation} 
where $R_0=c/H_0=9.25 \ h^{-1} \times 10^{27}$ cm, and $h=H_0/100$ is 
the normalized Hubble constant. Then, the observed peak flux is
\begin{equation} 
P=\frac{L_n}{4 \pi d_L^2} . 
\label{pf}
\end{equation}
Combining equations (\ref{ld}) and (\ref{pf}) yields 
\begin{equation}
1+z= \frac{1}{2} \left [1+0.3 h \sqrt{L_{n56}/P}+
\left (1+0.6 h \sqrt{L_{n56}/P} \right )^{1/2} \right ] ,
\end{equation}
where $L_{\rm n56}=L_{\rm n}/10^{56}$ is the normalized peak photon number luminosity.
Therefore, from equation (\ref{td}) we obtain
\begin{equation} 
\Dt= \frac{1}{2} \left [1+0.3 h \sqrt{L_{n56}/P}+
\left (1+0.6 h \sqrt{L_{n56}/P} \right )^{1/2} \right ] \Dt_0 ,
\end{equation}
which can be compared with observations.


\begin{thebibliography}{}
\bibitem[Balastegui et al. (2001)]{bala01} 
	Balastegui, A., Ruiz-Lapuente, P., \& Canal, R. 2001, MNRAS, 328, 283
\bibitem[Beloborodov et al. (1998)]{belo98}
	Beloborodov, A. M., Stern, B. E., \& Svensson, R. 1998, ApJ, 508, L25 
\bibitem[Beloborodov et al. (2000)]{belo00} 
	Beloborodov, A. M., Stern, B. E., \& Svensson, R. 2000, ApJ, 535, 158
\bibitem[Che et al. (1997)]{che97}
	Che, H., Yang, Y., Wu, M., \& Li, T. P. 1997, ApJ, 477, L69
\bibitem[Costa et al.(1997)]{costa97}
	Costa, E., et al. 1997, Nature, 387, 783
\bibitem[Kobayashi et al. (1997)]{koba97}
	Kobayashi, S., Piran, T., \& Sari, R. 1997, ApJ, 490, 92
\bibitem[Kouveliotou et al. (1993)]{kouve93}
	Kouveliotou, C., Meegan, C. A., Fishman, G. J., Bhat, N. P., Briggs, 
	M. S., Koshut, T. M., Paciesas, W. S., \& Pendleton, G. N. 1993, ApJ, 413, L101
\bibitem[Kulkarni et al.(1998)]{kulkarni98}
	Kulkarni, S. R., et al. 1998, Nature, 393, 35
\bibitem[Li (2001)]{li01}
	Li, T. P. 2001, Chin. J. Astron. Astrophys., 1, 313
\bibitem[Li, Muraki (2002)]{li02}
        Li, T. P., \& Muraki, Y. 2002, ApJ, 578, 374	
\bibitem[McBreen et al. (2001)]{mcbreen01}
	McBreen, S., Quilligan, F., McBreen, B., Hanlon, L., \& Watson, D. 
	2001, A\&A, 380, L31
\bibitem[Meegan et al.(1992)]{meegan92}
	Meegan, C. A., Fishman, G. J., Wilson, R. B., Paciesas, W. S., 
	Pendleton, G. N., Horack, J. M., Brock, M. N., \& Kouveliotou C. 
	1992, Nature, 355, 143
\bibitem[Metzger et al. (1997)]{metzger97}
	Metzger, M. R., Djorgovski, S. G., Kulkarni, S. R., Steidel, C. C., 
	Adelberger, K. L., Frail, D. A., Costa, E. \& Frontera, F. 
	1997, Nature, 387, 878
\bibitem[Mukherjee et al. (1998)]{mukherjee98}
	Mukherjee, S., Feigelson, E. D., Babu, G. J., Martagh, F., Fraley, C.,
	\& Raftery, A. 1998, ApJ, 508, 314
\bibitem[Nakar, Piran (2002)]{nakar02}
	Nakar, E., \& Piran, T. 2002, MNRAS, 331, 40
\bibitem[Norris et al. (1996)]{norris96}
	Norris, J. P., Nemiroff, R. J., Bonnell, J. T., Scargle, J. D., 
	Kouveliotou, C., Paciesas, W. S., Meegan, C. A., \& Fishman, G. J.
	1996, ApJ, 459, 393
\bibitem[Norris et al. (1994)]{norris94}
	Norris, J. P., Nemiroff, R. J., Scargle, J. D., Kouveliotou, 
	C., Fishman, G. J., Meegan, C. A., Paciesas, W. S., \& Bonnell, J. T. 
	1994, ApJ, 424, 540
\bibitem[Pozanenko, Loznikov (2000)]{pozanenko00}
 	Pozanenko, A. S., \& Loznikov, V. M. 2000, In: Gamma-Ray Bursts: 
	$5^{\rm th}$ Huntsville Symposium, AIP Conf. Proc. 526, R. M. Kippen, 
	R. S. Mallozzi, G. J. Fishman, ed. New York: AIP, 220
\bibitem[Piran (1999)]{piran99}
	Piran, T. 1999, Phys. Rep., 314, 575
\bibitem[Press et al. (1992)]{press92}
	Press, W. H., Teukolsky, S. A., Vetterling, W. T., \& Flannery, B. P. 
	1992, Numerical Recipes in FORTRAN, 2nd ed. 
	(Cambridge: Cambridge University Press), 614
\bibitem[Rees, M\'{e}sz\'{a}ros(1992)]{rees92}
	Rees, M. J. \& M\'{e}sz\'{a}ros, P. 1992, MNRAS, 258, 41
\bibitem[Rees, M\'{e}sz\'{a}ros(1994)]{rees94}
	Rees, M. J. \& M\'{e}sz\'{a}ros, P. 1994, ApJ, 430, L93
\bibitem[Sari, Piran (1997)]{sari97}
	Sari, R., \& Piran, T. 1997, ApJ, 485, 270			
\bibitem[van Paradijs et al. (1997)]{van97}
	van Paradijs, J., et al. 1997, Nature, 386, 686

\end{thebibliography}
\end{document}